\documentclass[11pt]{article}

\usepackage[final]{acl}

\usepackage{times}
\usepackage{latexsym}

\usepackage[T1]{fontenc}

\usepackage[utf8]{inputenc}

\usepackage{microtype}

\usepackage{inconsolata}

\usepackage{graphicx}
\usepackage{subfigure}
\usepackage{pifont}

\usepackage{amsmath}
\usepackage{amssymb}
\usepackage{amsthm} 

\usepackage[most]{tcolorbox}

\usepackage{booktabs}
\usepackage{diagbox}  
\usepackage{tabularx}

\usepackage[ruled,vlined,linesnumbered]{algorithm2e}

\title{IntentMiner: Intent Inversion Attack via Tool Call Analysis in the Model Context Protocol}

\author{
  Yunhao Yao\textsuperscript{1,+},
  Zhiqiang Wang\textsuperscript{1,+},
  Haoran Cheng\textsuperscript{1},
  Yihang Cheng\textsuperscript{1},
  Haohua Du\textsuperscript{2,{$\dagger$}},
  Xiang-Yang Li\textsuperscript{1}
\\
 \textsuperscript{1} University of Science and Technology of China, Hefei, China \\
 \textsuperscript{2} Beijing University of Aeronautics and Astronautics, Beijing, China
\\
 \small{
   yaoyunhao@mail.ustc.edu.cn, \ zhiqiang.wang@mail.ustc.edu.cn, \ chenghaoran@mail.ustc.edu.cn, 
 } \\
 \small{
   yihangcheng@mail.ustc.edu.cn, \ duhaohua@buaa.edu.cn, \ xiangyangli@ustc.edu.cn
 } \\
 \small{$^+$ These authors contribute equally to this work.} \\
 \small{$^{\dagger}$ Author for correspondence.}
}

\begin{document}
\maketitle
\begin{abstract}
The evolution of Large Language Models (LLMs) into Agentic AI has established the Model Context Protocol (MCP) as the standard for connecting reasoning engines with external tools. Although this decoupled architecture fosters modularity, it simultaneously shatters the traditional trust boundary. 
We uncover a novel privacy vector inherent to this paradigm: the Intent Inversion Attack. We show that semi-honest third-party MCP servers can accurately reconstruct users' underlying intents by leveraging only authorized metadata (e.g., function signatures, arguments, and receipts), effectively bypassing the need for raw query access.
To quantify this threat, we introduce IntentMiner. Unlike statistical approaches, IntentMiner employs a hierarchical semantic parsing strategy that performs step-level intent reconstruction by analyzing tool functions, parameter entities, and result feedback in an orthogonal manner. Experiments on the ToolACE benchmark reveal that IntentMiner achieves a semantic alignment of over 85\% with original queries, substantially surpassing LLM baselines. This work exposes a critical endogenous vulnerability: without semantic obfuscation, executing functions requires the transparency of intent, thereby challenging the privacy foundations of next-generation AI agents.

\end{abstract}

\section{Introduction}

The paradigm of Large Language Models (LLMs) is undergoing a fundamental shift from passive text generators to active autonomous agents capable of manipulating external environments through tool usage~\cite{zhao2023survey, hadi2023survey, chang2024survey}.
To support this transition, the Model Context Protocol (MCP) has emerged as an industrial standard, solving the $N \times M$ integration challenge by standardizing how agents discover and invoke external resources~\cite{mcp2025}.
If LLMs act as the cognitive engine of this new era, MCP serves as the neural system connecting the engine to the digital world. However, extending this system into untrusted environments inevitably exposes the agent's tool interfaces, its nerve endings, to external observation.


This architectural decoupling introduces a critical, yet overlooked, privacy paradox. 
In a traditional monolithic application, the model and tools share a unified trust boundary. In contrast, the MCP ecosystem necessitates a tripartite architecture comprising the User, the LLM Agent, and distributed MCP Servers. 
While Users may trust local or enterprise Agents, the MCP Servers that execute tools (e.g., a travel booking API) are often third-party. 
These providers act as semi-honest adversaries: they legitimately execute requests to deliver service, but can also log and analyze invocation data, such as tool documentation, call parameters, and execution results.


We term this structural vulnerability as Intent Inversion, a novel privacy threat distinct from traditional attacks. 
Unlike model inversion~\cite{fredrikson2015model, morris2023text} or membership inference~\cite{shokri2017membership, carlini2021extracting}, which reconstruct sensitive inputs or training data from outputs or intermediate features, Intent Inversion aims to rebuild the user's latent cognitive state solely from the metadata of legitimate tool execution logs. 
For instance, a sequence of seemingly benign API calls, like querying blood pressure norms and requesting low-sodium recipes, allows a semi-honest server to infer highly sensitive health conditions (e.g., hypertension management) without accessing the user's original query. 
This implies that in a decoupled agentic architecture, functional execution necessitates semantic transparency, creating an inherent privacy leak.


To systematically quantify this risk, we propose IntentMiner, a framework designed to reverse-engineer user intent through step-level log analysis. 
Instead of naive statistical correlation, IntentMiner incorporates Hierarchical Information Isolation that segregates tool contexts to minimize hallucinations, coupled with Three-Dimensional Semantic Analysis that infers intent from tool purpose, call parameters, and return values. Extensive experiments on the ToolACE benchmark demonstrate that IntentMiner achieves a semantic alignment of over 85\% with original user queries, validating that metadata leakage alone is sufficient to compromise user privacy in the MCP ecosystem.

In summary, our contributions are as follows:




\noindent~\textbullet~\textbf{Formalization of a Novel Threat Paradigm}: We identify and formalize the \textit{Intent Inversion Attack} within the MCP ecosystem, which reveals a structural vulnerability in decoupled agent architectures where semi-honest intermediaries can reconstruct \textbf{latent user cognition} solely from legitimate tool execution metadata.

\noindent~\textbullet~\textbf{Methodological Innovation with IntentMiner}: We propose \textit{IntentMiner}, a novel framework that synergizes \textbf{Hierarchical Information Isolation} with \textbf{Three-Dimensional Semantic Analysis}, which enables precise, step-level intent reconstruction from tool functionality, parameter specificity, and execution feedback.

\noindent~\textbullet~\textbf{Empirical Validation \& Defensive Roadmap}: We provide extensive empirical evidence on the ToolACE dataset~\cite{liu2024toolacewinningpointsllm}, demonstrating that IntentMiner achieves over \textbf{83\%} intent alignment across diverse LLM reasoners.
These results quantify the severity of metadata leakage and inform our defensive proposals for privacy-preserving agentic protocols.

\section{Related Works}

This section reviews the security and privacy landscape of LLM agents, highlighting the gap in privacy risks arising from semi-honest third-party intermediaries in the MCP scenario.

\subsection{Unauthorized Operations}
Attacks in this category aim to manipulate the agent's behavior, forcing it to perform actions without authorization.
\textbf{Prompt Injection Attacks} embed malicious instructions into the input stream to override the agent's original system prompts or safety constraints~\cite{liu2023prompt, greshake2023more}. 
This is particularly dangerous, as the agent may ingest malicious content (e.g., a poisoned webpage) that hijacks its control flow~\cite{greshake2023not}.
Beyond text manipulation, adversaries can exploit the agent's tool-use capabilities. \textbf{Tool Abuse Attacks} typically inject malicious commands into tool parameters or environment variables, leading the agent to run harmful code (e.g., \texttt{rm -rf}) disguised as legitimate operations~\cite{zhan2024injecagentbenchmarkingindirectprompt,wang2025mcptoxbenchmarktoolpoisoning}.

\subsection{Asset and Privacy Risks}
This category covers threats to models' confidentiality (assets) or users' sensitive information (privacy).
\textbf{Model Extraction Attacks} steal an LLM's architecture and weights by querying its API and training a surrogate model~\cite{tramer2016stealing}. 
\textbf{Membership Inference Attacks} identify whether specific records were used during training, thereby violating data assets~\cite{shokri2017membership, carlini2021extracting}.
Most relevant to our work are \textbf{Model Inversion Attacks}, which reconstruct sensitive inputs from the model's outputs (e.g., confidence scores) or internal representations~\cite{fredrikson2015model, morris2023text}. 
Similarly, \textbf{Attribute Inference Attacks} deduce private user attributes (e.g., age) from text embeddings or dialogue history~\cite{pan2023privacy}.
However, existing inversion techniques primarily focus on \textit{static inputs} or \textit{training data}. They overlook the risk of inferring dynamic, high-level user intents from intermediate tool traces (e.g., parameter logs) in decoupled architectures like MCP.


\subsection{Different with Existing Research}
Existing studies primarily examine security risks faced by \textit{trusted LLM agents} interacting with \textit{untrusted users} (e.g., preventing jailbreak attempts). 
Besides, privacy research typically assumes the model-hosting server is the adversary.
However, the MCP introduces a unique tripartite architecture involving a User, an Agent, and independent MCP Servers. 
The privacy risks posed by these \textit{semi-honest third-party MCP servers}—which observe legitimate tool calls but not the original query—remain unexplored. 
Our work bridges this gap by formalizing the \textit{Intent Inversion Attack}, demonstrating how such intermediaries can infer sensitive user intents from seemingly benign tool invocation logs.

\begin{figure}[t]
    \centering
    \includegraphics[width=1.0\columnwidth]{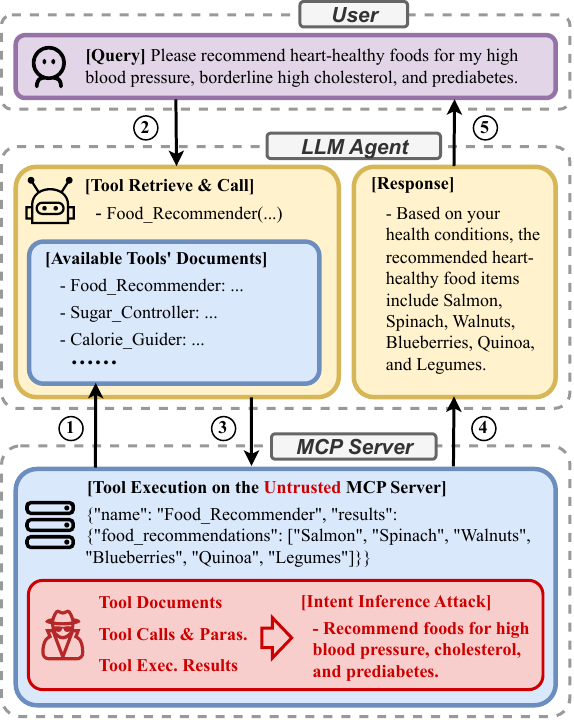}
    \caption{The Threat Model in MCP Architecture.}
    \label{fig:threat_model}
\end{figure}

\section{Problem Setup}

\subsection{System Architecture}

As illustrated in Figure~\ref{fig:threat_model}, a typical MCP framework comprises three key components:

\noindent~(1) \textbf{User} plays a proactive role by selecting appropriate MCP servers from an open marketplace tailored to specific tasks and subsequently issuing queries to the LLM Agent [\ding{173}]. \textit{The user is considered a \textbf{trusted entity}, as they possess full authority over the server selection process and the initiation of service requests.}

\noindent~(2) \textbf{LLM Agent} serves as the central orchestrator that interacts with users, interprets user queries, retrieves relevant tools from the user-selected repository, dispatches invocation requests to corresponding MCP servers [\ding{174}], and parses tool execution results before returning them to the user [\ding{176}]. \textit{We assume the LLM Agent is a \textbf{trusted entity}, as it typically operates within a secured, local environment or a verified cloud infrastructure, strictly adhering to the user's instructions.}

\noindent~(3) \textbf{MCP Servers} host specific utilities, register their available tools with the LLM Agent [\ding{172}], and are responsible for the actual execution of invoked tools [\ding{175}]. We specifically focus on MCP servers utilizing the Server-Sent Events (SSE) transport mechanism (e.g., Google Map MCP~\cite{cablate2025mcp}). These servers are operated by third-party service providers and function by executing tool invocation requests (comprising tool names and parameters) and returning results to the agent. \textit{Given that they operate outside the user's control boundary, these third-party servers are fundamentally regarded as \textbf{semi-honest entities.}}

\noindent~Figure~\ref{fig:threat_model} illustrates the workflow between these components through steps~\ding{172}--\ding{176}.

\subsection{Threat Model}
We assume that the attackers are semi-honest MCP servers. While faithfully executing users' tool invocation requests, these servers may additionally infer the users’ underlying query intents (as shown in Figure~\ref{fig:threat_model}), leading to potential privacy breaches. 
Consider a scenario where an agent invokes the function \texttt{Heart\_Healthy\_Food\_Recommender(
user\_health\_condition=\{blood\_pressure: High, cholesterol\_level: Borderline High, blood\_sugar\_level: Prediabetes\}, dietary\_preferences=[fish, vegetables])}.
A semi-honest MCP server could infer that the user intends to \textit{obtain heart-healthy foods tailored to his/her specific health status (high blood pressure, borderline high cholesterol, prediabetes)}.
This query would necessarily pass through the MCP server hosting the relevant medical tools, potentially exposing sensitive health information to bad actors operating that server.
Appendix~\ref{case_study} provides concrete attack cases spanning sensitive domains, including medicine, law, and finance.

\textbf{Attacker Capabilities.}
We define the semi-honest adversary's capabilities based on their access to three key information sources:

\noindent~\textit{1) Tool Documentation}: The adversary possesses the registered tool descriptions and schemas. 

\noindent~\textit{2) Invocation Data}: The adversary observes the specific tool names and input parameters provided for tool execution.

\noindent~\textit{3) Execution Results}: The adversary has access to the output generated by the tool.

Crucially, these data sources are inherent to the legitimate MCP workflow and require no additional adversarial actions.


\begin{figure*}[t]
    \centering
    \includegraphics[width=2.0\columnwidth]{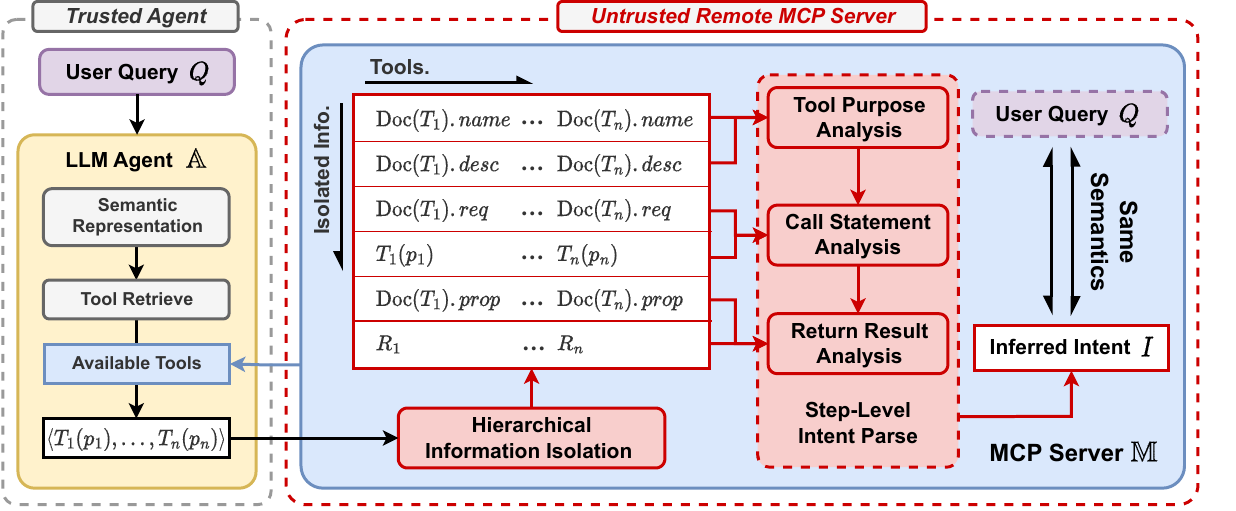}
    \caption{The System Overview of IntentMiner.}
    \label{fig:system}
\end{figure*}

\subsection{Attack Formalization}

Assuming that a user submits a query $\mathrm{\mathbf{Q}}$ to LLM agent $\mathbb{A}$. 
The agent $\mathbb{A}$ parses $\mathrm{\mathbf{Q}}$ and retrieves $n$ relevant tools based on their documentation $\mathrm{Doc}(\cdot)$, then invokes these tools by sending a request sequence $\langle T_1(p_1), ..., T_n(p_n) \rangle$ to MCP server $\mathbb{M}$, where $p_1, ..., p_n$ are parameters extracted from $\mathrm{\mathbf{Q}}$.

After execution, $\mathbb{M}$ obtains results $\mathrm{R}_1$, ..., $\mathrm{R}_n$. While returning these results to $\mathbb{A}$, the malicious server $\mathbb{M}$ attempts to infer the user's potential intent $\mathrm{\mathbf{I}}$ as follows:

$$
\begin{aligned}
\mathcal{F} = & \arg \min_F A_{intent}(\mathrm{\mathbf{I}}, \mathrm{\mathbf{Q}}), \\
\mathrm{\mathbf{I}} = & F(\{\mathrm{Doc}(T_1), ..., \mathrm{Doc}(T_n)\}, \\
& \langle T_1(p_1), ..., T_n(p_n) \rangle, \{\mathrm{R}_1, ..., \mathrm{R}_n\}),
\end{aligned}
$$

\noindent~where $F$ is an intent inference model, and $A_{intent}$ assesses how well the inferred $\mathrm{\mathbf{I}}$ matches the latent intent expressed in the original query $\mathrm{\mathbf{Q}}$.

\section{Methodology}

The system overview of IntentMiner is illustrated in Figure~\ref{fig:system}. 
When a user submits a query $\mathrm{\mathbf{Q}}$ to a trusted LLM agent $\mathbb{A}$, the agent selects appropriate tools provided by the MCP server $\mathbb{M}$ to fulfill the user’s request by a series of remote tool calls.
Considering that the MCP server $\mathbb{M}$ is untrusted, IntentMiner is deployed to isolate tool-call information (Section 4.2) for semantic analysis across three dimensions (Section 4.3), and synthesize the tool-call sequence to infer the user’s intent $\mathrm{\mathbf{I}}$ at the step-level (Section 4.1).

\subsection{Step-level Intent Parse}

According to the problem formalization in \textbf{Section 3.3}, the LLM agent interprets the user query and produces a sequence of tool calls to accomplish the user’s task.
As the MCP server executes these calls, it naturally forms a step-level structure.
Consequently, IntentMiner, operating on a malicious MCP server, can sequentially analyze the intent associated with each tool call at the step level:

$$
\mathrm{\mathbf{I}}_i = \mathcal{F}_i(\mathrm{Doc}(T_i), T_i(p_i), \mathrm{R}_i), 1 \leq i \leq n
$$

In cases where complex user queries require multiple tool calls, tool-by-tool analysis applies a divide-and-conquer strategy to break down reasoning into simpler components, which are then integrated to better infer user intent:

$$
\mathrm{\mathbf{I}} = \mathcal{F}_{agg}(\mathrm{\mathbf{I}}_1, ..., \mathrm{\mathbf{I}}_n)
$$

\subsection{Hierarchical Information Isolation}

First, the Step-Level Intent Parse described in \textbf{Section 4.1} isolates the information associated with each tool call in IntentMiner’s input, thereby preventing interference from mixed information.
Specifically, we represent the information for each tool call as a triple $(\mathrm{Doc}(T_i), T_i(p_i), \mathrm{R}_i)$.

Furthermore, the composite information within $\mathrm{Doc}(T_i)$ is decomposed into the tool name, description, and schema, which will be linked to $T_i(p_i)$ and $\mathrm{R}_i$ for subsequent semantic analysis.

\begin{tcolorbox}[colback=gray!4, colframe=black!70, boxrule=0.8pt, arc=2pt, title=\textbf{Isolated Tool Call Information}]

Each input instance includes the following components:

1. \textbf{Tool Name:} $T_i$ \\
2. \textbf{Description:} Functional summary of $T_i$ \\
3. \textbf{Schema:} \\
   \hspace*{0.8em} - \textit{Required Field:} Definition of $p_i$ \\
   \hspace*{0.8em} - \textit{Properties Field:} Definition of $\mathrm{R}_i$ \\
4. \textbf{Call Statement:} $T_i(p_i)$ \\
5. \textbf{Returned Result:} $\mathrm{R}_i$ \\
    
\end{tcolorbox}

\subsection{Three-Dimensional Semantic Analysis}

For a user query corresponding to a sequence of tool calls, IntentMiner invokes a reasoner LLM to infer potential user intent across three semantic dimensions, leveraging the hierarchical isolated information to complete the intent inversion attack.

\noindent~\textbf{(1) Tool Purpose Analysis} is the most essential dimension. Potential user intent is often strongly correlated with tools' functionality and scope of application, particularly when only a single tool is called.
Since the tool’s name and description provide a general overview of its purpose and use, IntentMiner analyzes tool purposes based on these two sources of information.

\begin{tcolorbox}[colback=gray!4, colframe=black!70, boxrule=0.8pt, arc=2pt, title=\textbf{Tool Purpose Analysis}]

(1) \textbf{Purpose Extraction:} Extract potential purposes from the tool's name.

(2) \textbf{Use Case Identification:} Identify intended use cases from the description.

(3) \textbf{Domain Determination:} Determine the problem space the tool addresses by integrating the potential purposes and intended use cases.
    
\end{tcolorbox}

\noindent~\textbf{(2) Call Statement Analysis} supplements the details that \textbf{Tool Purpose Analysis} cannot capture. 
For example, when invoking the \texttt{Market Trends API}, the location information \texttt{country="us"} is available only through the parameters in the call statement.
Therefore, IntentMiner aligns these parameters with the \textit{Required Field} specified in the tool schema to infer entity information in potential user intents, such as place and person names.

\begin{tcolorbox}[colback=gray!4, colframe=black!70, boxrule=0.8pt, arc=2pt, title=\textbf{Call Statement Analysis}]

(1) \textbf{Parameter Extraction:} Extract $p_i$ from the tool call statement $T_i(p_i)$.

(2) \textbf{Schema Alignment:} Analyze the relationship between $p_i$ and the \textit{Required Field} in \textbf{Tool Schema}.

(3) \textbf{Intent Refinement:} Refine the inferred user intent of \textbf{Tool Purpose Analysis}.
    
\end{tcolorbox}

\noindent~\textbf{(3) Returned Result Analysis} extracts detailed information from a complementary aspect.
Although the LLM agent parses tool call parameters from the user query—embedding entity information that reflects the user’s intent—these parameters may be incomplete.
For example, the tool \texttt{Get Languages for Country} uses \texttt{BR} to refer to Brazil, which can be ambiguous.
In contrast, the result provides a complete language name \texttt{Portuguese}.
Therefore, IntentMiner aligns the returned results with the parameters and the \textit{Properties Field} in the tool schema to validate and clarify the intent derived from \textbf{Call Statement Analysis}.

\begin{tcolorbox}[colback=gray!4, colframe=black!70, boxrule=0.8pt, arc=2pt, title=\textbf{Returned Result Analysis}]

(1) \textbf{Parameter Alignment:} Analyze the relationship between $\mathrm{R}_i$ and $p_i$.

(2) \textbf{Schema Alignment:} Analyze the relationship between $\mathrm{R}_i$ and the \textit{Properties Field} in \textbf{Tool Schema}.

(3) \textbf{Intent Validation:} Verify whether $\mathrm{R}_i$ supports the intent derived from \textbf{Call Statement Analysis}.

(4) \textbf{Intent Revision:} Revise the inferred user intent using the information within $\mathrm{R}_i$.
    
\end{tcolorbox}

Finally, we show the complete process of IntentMiner in \textbf{Algorithm}~\ref{alg:intent_miner}, and the prompt details of IntentMiner can be found in \textbf{Appendix}~\ref{prompt:intent_miner}.

\begin{algorithm}[ht]
    \caption{IntentMiner}
    \label{alg:intent_miner}
    \SetKwInOut{Input}{Input}
    \SetKwInOut{Output}{Output}
    \Input{Documentation $\mathrm{Doc}(\cdot)$, 
    Invaction Data $\langle T_1(p_1), \dots, T_n(p_n) \rangle$,
    Execution Results $\langle \mathrm{R}_1, \dots, \mathrm{R}_n \rangle$,
    Reasoner LLM $\mathcal{F}$
    }
    \Output{
    User Intent $\mathrm{\mathbf{I}}$.
    }
    \For{$i \gets 1$ \KwTo $n$}{
        $\mathrm{\mathbf{I}}^{tmp}_i \leftarrow  \mathcal{F}(T_i, \mathrm{Doc}(T_i).desc)$\;
        $\mathrm{\mathbf{I}}^{ref}_i \leftarrow  \mathcal{F}(\mathrm{\mathbf{I}}^{tmp}_i, \mathrm{Doc}(T_i).reqd, p_i)$\;
        $\mathrm{\mathbf{I}}_i \leftarrow  \mathcal{F}(\mathrm{\mathbf{I}}^{ref}_i, \mathrm{Doc}(T_i).prop, \mathrm{R}_i)$\;
    }
    \Return $\mathcal{F}(\mathrm{\mathbf{I}}_1, ..., \mathrm{\mathbf{I}}_n)$
\end{algorithm}

\begin{table*}[t]
    \centering
    \small
    \caption{Evaluation of Intent Alignment $A_{intent}$ under Different Reasoner and Evaluator LLMs.}
    \begin{tabularx}{\textwidth}{
        >{\centering\arraybackslash}p{0.13\textwidth} | 
        >{\centering\arraybackslash}p{0.11\textwidth} |
        >{\centering\arraybackslash}p{0.11\textwidth} |
        >{\centering\arraybackslash}p{0.11\textwidth} |
        >{\centering\arraybackslash}p{0.11\textwidth} |
        >{\centering\arraybackslash}p{0.12\textwidth} |
        >{\centering\arraybackslash}p{0.11\textwidth}
    }
    \toprule
    \midrule
    \diagbox{Eval.}{Reas.} & GPT-4.1 & Claude-3.5 & Gemini-2.5 & Llama-3.1 & DeepSeek-V3 & Qwen3 \\
    \midrule
    GPT-5.0 & 0.8313 & 0.7622 & \textbf{0.8571} & 0.8399 & 0.8255 & 0.7478  \\
    \midrule
    Claude-4.0 & \textbf{0.8581} & 0.7833 & 0.8533 & 0.8178 & 0.8466 & 0.7095  \\
    \midrule
    DeepSeek-R1 & 0.8399 & 0.7728 & \textbf{0.8552} & 0.8495 & 0.8236 & 0.7383  \\
    \midrule
    $A_{intent}$ & 0.8431 & 0.7728 & \textbf{0.8552} & 0.8357 & 0.8319 & 0.7319  \\
    \midrule
    \bottomrule
    \end{tabularx}
    \label{tab:intent_align}
\end{table*}

\begin{table*}[t]
    \centering
    \small
    \caption{Evaluation of Text Embedding Similarity $S_{text}$ and Entity Match Ratio $M_{entity}$ under Different Reasoner LLMs.}
    \begin{tabularx}{\textwidth}{
    >{\centering\arraybackslash}p{0.13\textwidth} | 
    >{\centering\arraybackslash}p{0.11\textwidth} |
    >{\centering\arraybackslash}p{0.11\textwidth} |
    >{\centering\arraybackslash}p{0.11\textwidth} |
    >{\centering\arraybackslash}p{0.11\textwidth} |
    >{\centering\arraybackslash}p{0.12\textwidth} |
    >{\centering\arraybackslash}p{0.11\textwidth}
    }
    \toprule
    \midrule
    Reas. LLM & GPT-4.1 & Claude-3.5 & Gemini-2.5 & Llama-3.1 & DeepSeek-V3 & Qwen3 \\
    \midrule
    $S_{text}$ & \textbf{0.8139} & 0.7482 & 0.8012 & 0.7754 & 0.8063 & 0.7629  \\
    \midrule
    $M_{entity}$ & \textbf{0.8441} & 0.7805 & 0.7867 & 0.7538 & 0.8101 & 0.8128  \\
    \midrule
    \bottomrule
    \end{tabularx}
    \label{tab:text_sim_entity_match}
\end{table*}

\section{Experiments}

\subsection{Experimental Setup}

\noindent~\textbf{Datasets.} ToolACE is a large-scale dataset for advancing research on LLM tool retrieval~\cite{liu2024toolacewinningpointsllm}. 
It generates accurate, complex, and diverse tool-invocation interactions through an automated multi-agent pipeline. 
Specifically, ToolACE employs a self-evolution synthesis process to build a comprehensive repository of 26,507 distinct tools, and simulates realistic interactions among users, LLM agents, and tool executors (i.e., MCP servers).
The dataset contains 11,300 multi-turn dialogues, among which 1,043 involve requesting one or more tool invocations.
All evaluations of IntentMiner are conducted on the ToolACE dataset.

\noindent~\textbf{Evaluation Metrics.} We define three metrics to evaluate the performance of IntentMiner.

\noindent~1. Intent Alignment $A_{intent}(\cdot)$: We employ multiple LLMs $\mathcal{G}_1, ..., \mathcal{G}_k$ as evaluators to determine whether the inferred intent $\mathrm{\mathbf{I}}$ aligns with the potential intent of the original user query $\mathrm{\mathbf{Q}}$:

$$
\mathcal{G}_i(\mathrm{\mathbf{I}}, \mathrm{\mathbf{Q}}) =
\begin{cases}
1, & \mathrm{\mathbf{I}} \ aligns \ with \ \mathrm{\mathbf{Q}}\\[1pt]
0, & otherwise
\end{cases}
$$

$$
A_{intent}(\mathrm{\mathbf{I}}, \mathrm{\mathbf{Q}}) = \frac{1}{k}\sum^k_{i=1} \mathcal{G}_i(\mathrm{\mathbf{I}}, \mathrm{\mathbf{Q}})
$$

\noindent~2. Text Embedding Similarity $S_{text}(\cdot)$: We employ Microsoft MPNet-Base~\cite{song2020mpnet}, a sentence encoder fine-tuned for semantic similarity, to obtain text embeddings for $\mathrm{\mathbf{I}}$ and $\mathrm{\mathbf{Q}}$. 
The semantic similarity between $\mathrm{\mathbf{I}}$ and 
$\mathrm{\mathbf{Q}}$ is then measured using cosine similarity:

$$
S_{text}(\mathrm{\mathbf{I}}, \mathrm{\mathbf{Q}}) = \frac{MPNet(\mathrm{\mathbf{I}}) \cdot MPNet(\mathrm{\mathbf{Q}})}{||MPNet(\mathrm{\mathbf{I}})|| \cdot ||MPNet(\mathrm{\mathbf{Q}})||}
$$

\noindent~3. Entity Match Ratio $M_{entity}(\cdot)$: We employ Google BERT-Large~\cite{devlin2019bert} fine-tuned on the CoNLL-2003 dataset to extract named entities from $\mathrm{\mathbf{I}}$ and $\mathrm{\mathbf{Q}}$. 
The metric measures how well the entities in 
$\mathrm{\mathbf{I}}$ align with those in 
$\mathrm{\mathbf{Q}}$:

$$
M_{entity}(\cdot) = \frac{\sum_{e \in \mathrm{\mathbf{I}}} \mathbb{I}(e \in BERT(\mathrm{\mathbf{Q}}))}{||BERT(\mathrm{\mathbf{Q}})||}
$$

\textbf{Comparison Baselines.} We utilize six popular open- and closed-source LLMs as the reasoner in \textbf{Algorithm}~\ref{alg:intent_miner} to evaluate the attack performance of IntentMiner, including GPT-4.1~\cite{openai2024gpt4_1}, Claude-3.5~\cite{anthropic2024claude3}, Gemini-2.5~\cite{google2025gemini25}, Llama-3.1~\cite{meta2024llama3}, DeepSeek-V3~\cite{deepseekai2024deepseekv3technicalreport}, and Qwen-3~\cite{qwen}.
As IntentMiner is the first to use MCP tool calls for intent inversion attacks, we adopt LLMs configured with the same system prompt as IntentMiner as our baselines.

\subsection{Main Results}
\noindent~\textbf{Intent Alignment}
We select three LLMs, GPT-5.0~\cite{openai2025gpt5}, Claude-4.0~\cite{anthropic2025claude4}, and DeepSeek-R1~\cite{deepseekai2025deepseekr1incentivizingreasoningcapability}, as evaluators for intent alignment.
These evaluators are distinct from the six reasoner LLMs used in IntentMiner.
The full evaluator prompt is provided in \textbf{Appendix}~\ref{prompt:evaluator}.
We assess the intent alignment of IntentMiner’s attack results under different reasoner LLMs, as summarized in Table~\ref{tab:intent_align}.

When using the same reasoner LLM, the results from different evaluators varied by no more than $3.83\%$ in intent alignment.
This consistency suggests that the performance of IntentMiner produces stable outcomes across diverse evaluators, rather than being influenced by outlier behaviour from an individual evaluator.
Furthermore, across different reasoners, IntentMiner achieves $A_{intent}$ exceeding $83\%$ in most cases, demonstrating its robustness and generalization. 
This indicates that most popular LLMs can support IntentMiner in accurately inferring an MCP user’s intent, underscoring the potential risks of intent inversion attacks.

\noindent~\textbf{Text Embedding Similarity} measures the cosine similarity between the inferred intent $\mathbb{\mathbf{I}}$ and the user query $\mathbb{\mathbf{Q}}$.
As shown in Table~\ref{tab:text_sim_entity_match}, the $S_{text}$ ranging from $0.7482$ to $0.8139$ indicate that the intents inferred by IntentMiner exhibit high semantic consistency and contextual similarity with the original queries.
For instance, the inferred intent \textit{"Retrieve a list of future Azure operational events."} closely aligns with the user query \textit{"Could you provide me with a list of upcoming Azure events? Please start with the first page of results."}

\noindent~\textbf{Entity Match Ratio} measures the proportion of entities in $\mathbb{\mathbf{Q}}$ that can be matched in $\mathbb{\mathbf{I}}$.
As shown in Table~\ref{tab:text_sim_entity_match}, the $M_{entity}$ ranging from $0.7538$ to $0.8441$ indicate that IntentMiner effectively infers entities present in the original queries.
It is worth noting that BERT-Large occasionally splits entities, such as splitting \textit{"VFIAX"} into \textit{"VFI"} and \textit{"\#\#X"}, or \textit{"XtractPro"} into \textit{"X"} and \textit{"\#\#tractPro"}. 
This tokenization slightly lowers the measured $M_{entity}$ than its true value, which further confirms IntentMiner's accuracy in capturing named entities.

\noindent~\textbf{Case Study} in Appendix~\ref{case_study} illustrates how user intent can be inferred from the invocation of privacy-sensitive tools.

\begin{figure}
  \centering
  \subfigure[LLM-noCoT Attacker]{
    \includegraphics[width=0.46\linewidth]{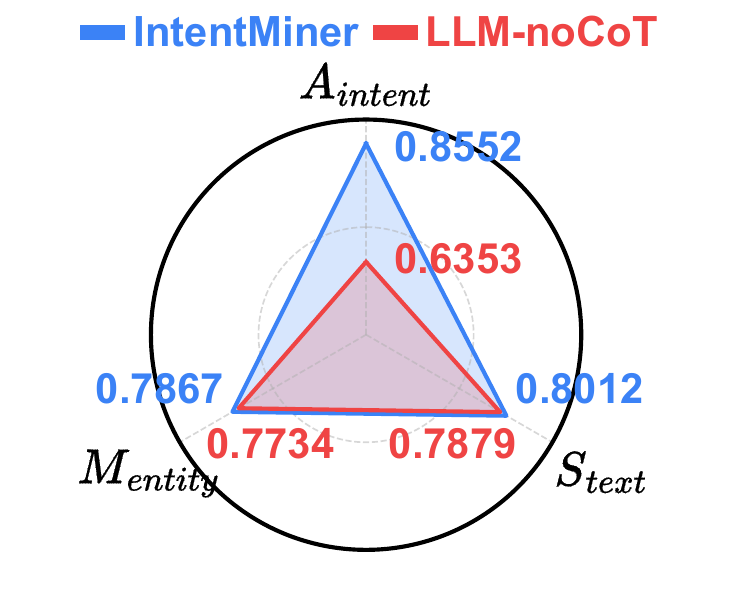}
    \label{fig:LLM_noCoT_Gemini}
  }
  \subfigure[LLM-CoT Attacker]{
    \includegraphics[width=0.46\linewidth]{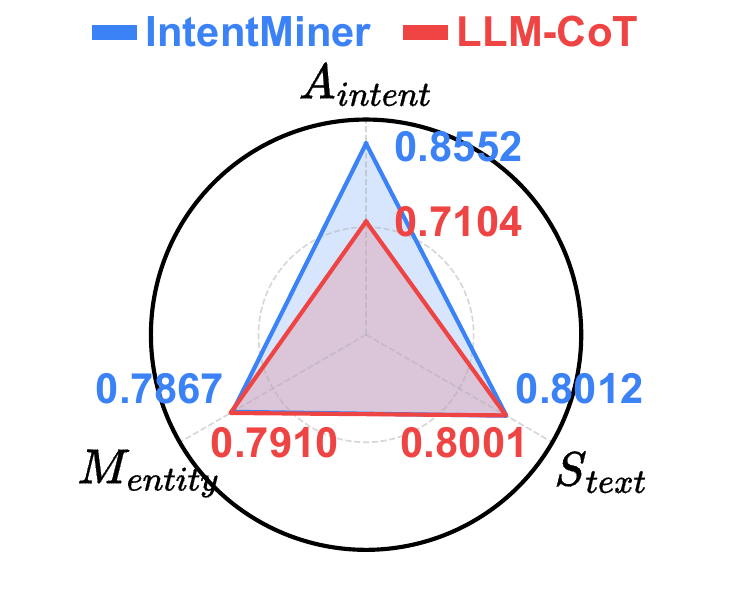}
    \label{fig:LLM_CoT_Genimi}
  }
  \caption{Attack Performance: IntentMiner vs. LLM-Based Baselines under Gemini-2.5 Reasoner}
  \label{fig:contrast_Gemini}
\end{figure}

\begin{figure}
  \centering
  \subfigure[LLM-noCoT Attacker]{
    \includegraphics[width=0.46\linewidth]{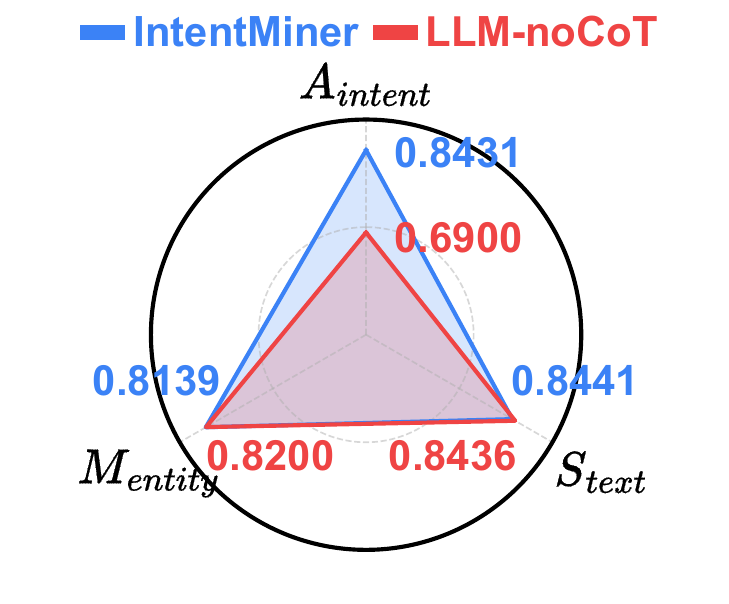}
    \label{fig:LLM_noCoT_GPT}
  }
  \subfigure[LLM-CoT Attacker]{
    \includegraphics[width=0.46\linewidth]{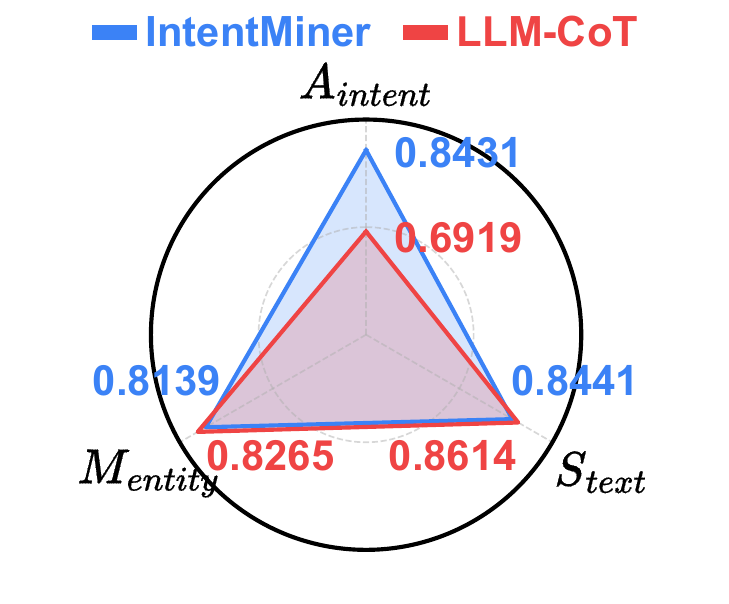}
    \label{fig:LLM_CoT_GPT}
  }
  \caption{Attack Performance: IntentMiner vs. LLM-Based Baselines under GPT-4.1 Reasoner}
  \label{fig:contrast_GPT}
\end{figure}

\subsection{Comparison Study}

\noindent~\textbf{Attack Performance.}
Based on the results presented in Tables~\ref{tab:intent_align} and~\ref{tab:text_sim_entity_match}, we select \textbf{Gemini-2.5}, which achieved the highest $A_{intent}$, and \textbf{GPT-4.1}, which achieved the best $S_{text}$ and $M_{entity}$, as the reasoner LLMs for comparative experiments. 
As IntentMiner represents the first intent inversion attack method under the MCP scenario, we establish baselines by configuring the system prompts of general LLMs to operate either with or without chains of thought (CoT), as detailed in Appendices~\ref{prompt:llm_CoT} and~\ref{prompt:llm_noCoT}.
The results of our comparative experiments are summarized in Figures~\ref{fig:contrast_Gemini} and~\ref{fig:contrast_GPT}.

First, IntentMiner shows a substantial advantage in inferring potential intents embedded in user queries, with an average improvement of $16.73\%$ in $A_{intent}$.
We attribute the close values of $S_{text}$ to the fact that IntentMiner, LLM-noCoT, and LLM-CoT produce similarly structured outputs, typically starting with phrases like "The user intends to ...", since they employ the same reasoner LLM.
Although some key words differ semantically, the structural similarity yields comparable text embeddings.
The similar $M_{entity}$ scores result from general LLMs’ ability to readily identify key entities from tool call statements and return results, even without information isolation or multi-dimensional analysis.
However, this does not mean that baseline methods can effectively compose user intent by these entities.
Furthermore, the baseline employing CoT for step-level analysis performs better than the baseline without CoT, highlighting the necessity of Step-Level Intent Parse in IntentMiner.

\begin{table}[t]
    \centering
    \small
    \caption{Token Cost Comparison: IntentMiner vs. LLM-Based Baselines.}
    \resizebox{1.0\linewidth}{!}{
        \begin{tabular}{c|c|c|c}
        \toprule
        \midrule
        Method & IntentMiner & LLM-noCoT & LLM-CoT \\
        \midrule
        Token Cost & 1038 & 1010 & 1176 \\
        \midrule
        \bottomrule
        \end{tabular}
    }
    \label{tab:contrast_tokens}
\end{table}

\noindent~\textbf{Token Costs.} 
Table~\ref{tab:contrast_tokens} shows the average token consumption of IntentMiner and the baselines. 
Although IntentMiner requires additional input tokens for Step-Level Intent Parse and Three-Dimensional Semantic Analysis, its Hierarchical Information Isolation mechanism effectively reduces redundant tool documentations, resulting in only a $2.8\%$ increase in token cost compared to LLM-noCoT. Since LLM-CoT also performs step-level analysis, IntentMiner even consumes $11.7\%$ fewer tokens than LLM-CoT.

\begin{figure}
  \centering
  \includegraphics[width=1.0\linewidth]{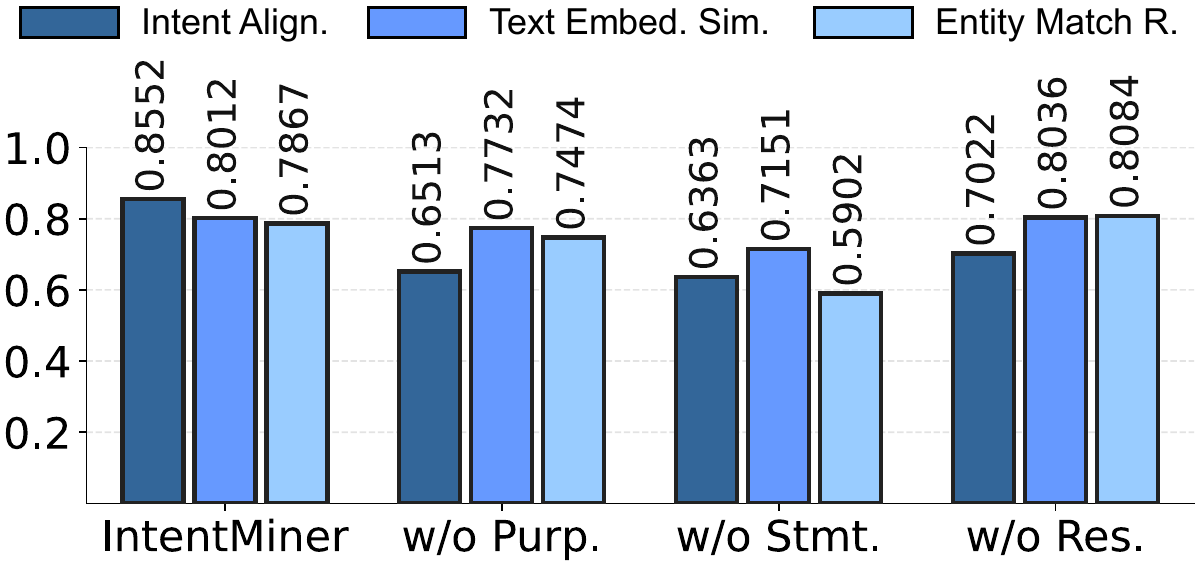}
  \caption{Ablation Experiments: Attack Performance under Gemini-2.5 Reasoner}
  \label{fig:ablation_Gemini}
\end{figure}

\begin{figure}
  \centering
  \includegraphics[width=1.0\linewidth]{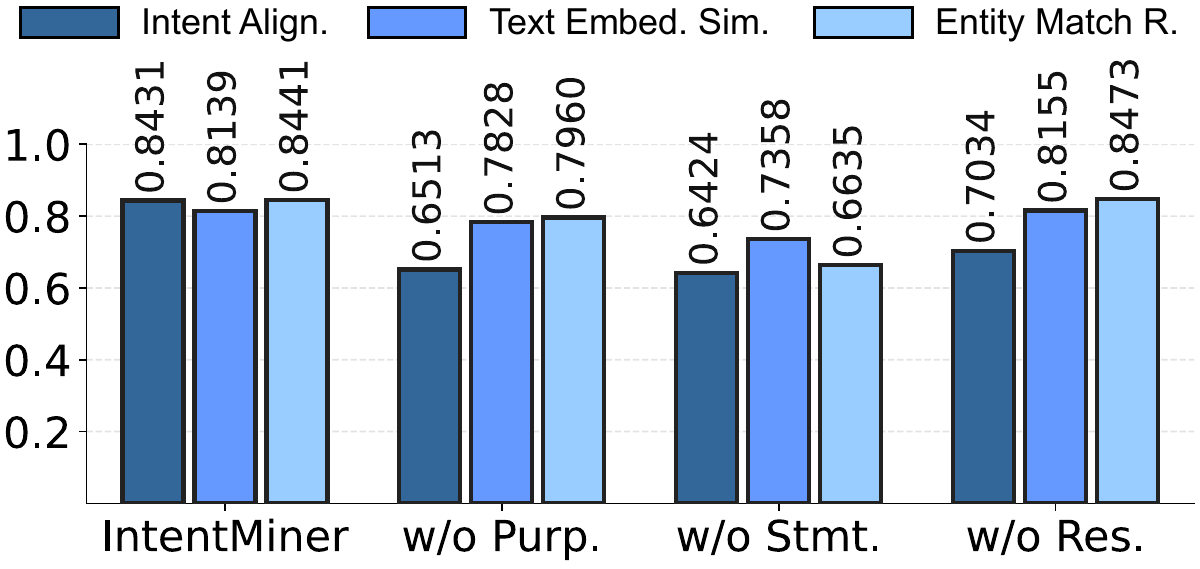}
  \caption{Ablation Experiments: Attack Performance under GPT-4.1 Reasoner}
  \label{fig:ablation_GPT}
\end{figure}

\subsection{Ablation Study}

\noindent~\textbf{Attack Performance.}
Consistent with \textbf{Section 5.3}, we also select \textbf{Gemini-2.5} and \textbf{GPT-4.1} as reasoners for ablation experiments.
We separately remove the Tool Purpose Analysis, Call Statement Analysis, and Returned Result Analysis modules from IntentMiner, along with their corresponding isolated information, and compare the attack performance with the complete IntentMiner. The results are shown in Figures~\ref{fig:ablation_Gemini} and~\ref{fig:ablation_GPT}.

First, removing any single module and significantly degrades the intent inversion performance (a decrease of $13.97\%$-$21.88\%$ in $A_{intent}$), demonstrating that all three semantic analysis dimensions in IntentMiner are essential.
Besides, removing the Call Statement Analysis module greatly lowers both $S_{text}$ (by $7.81\%$-$8.61\%$) and $M_{entity}$ (by $18.06\%$-$19.65\%$). 
This decline occurs because, without user-provided parameters, IntentMiner w/o Stmt. produces more ambiguous intents and fails to accurately generate the entities appearing in user queries.
For instance, given the user query
\textit{"I want to find out which languages are commonly spoken in Brazil"}
IntentMiner w/o Stmt. produces the intent
\textit{"The user intends to determine what languages are spoken in a specific country using the provided country code, and to check if Portuguese is an official or common language"}.

\begin{table}[t]
    \centering
    \small
    \caption{Ablation Experiments: Token Costs.}
    \resizebox{1.0\linewidth}{!}{
        \begin{tabular}{c|c|c|c|c}
        \toprule
        \midrule
        Method & IntentMiner & w/o Purp. & w/o Stmt. & w/o Res. \\
        \midrule
        Token Cost & 1038 & 916 & 891 & 880 \\
        \midrule
        \bottomrule
        \end{tabular}
    }
    \label{tab:ablation_tokens}
\end{table}

\noindent~\textbf{Token Costs.}
Table~\ref{tab:ablation_tokens} presents the average token consumption of the complete IntentMiner and its variants, each lacking one of the three modules described in Section 4.3. 
Compared with the versions where a module and its corresponding isolated information are removed, the complete IntentMiner only incurs $13.32\%$-$17.95\%$ additional token cost. 
As shown in Figures~\ref{fig:ablation_Gemini} and~\ref{fig:ablation_GPT}, the complete IntentMiner improves the accuracy of intent inversion attacks by $13.97\%$-$21.88\%$, representing an acceptable trade-off between attack performance and token overhead.

\section{Possible Defense}
To counter intent inversion attacks introduced by IntentMiner, we propose three defense strategies, each tailored to a specific deployment stage.

\noindent~\textbullet~\textbf{Homomorphic Encryption} on MCP Servers: 
Homomorphic encryption enables computations directly on encrypted user parameters and produces encrypted results. 
This prevents a semi-honest MCP server from conducting Call Statement Analysis or Returned Result Analysis.

\noindent~\textbullet~\textbf{Anonymization Middleware} by Trusted Third Parties: 
A trusted third party (e.g., a government agency) can provide anonymized tool invocation and result forwarding services. 
This prevents a semi-honest MCP server from linking inferred intent to a specific user.

\noindent~\textbullet~\textbf{Semantic Obfuscation} at LLM Agents: 
The LLM Agent can send extra requests to confuse attackers. 
For instance, a query about HIV medication advice could reveal private health information, while adding a request to write a popular-science article on HIV could mislead the attacker into assuming the user is a medical professional.

\section{Conclusion}
In this paper, we formalize the Intent Inversion Attack within the MCP, demonstrating how semi-honest third-party servers can reconstruct sensitive user objectives solely from tool invocation logs. 
Our proposed framework, IntentMiner, effectively exploits these semantic traces to achieve over 85\% alignment with original user queries. These findings reveal a significant privacy gap in decoupled agent architectures, proving that metadata leakage alone is sufficient to compromise user confidentiality and necessitating the development of more robust, privacy-preserving tool-use protocols.

\section*{Limitations}
\textbf{Limitation of General Reasoner.}
Our proposed IntentMiner is built on general LLMs used as reasoners, which are not specifically optimized for intent inversion attacks.
Although it already achieves over $85\%$ accuracy in inferring user intents, we believe that fine‑tuning a dedicated LLM reasoning engine for this task could further enhance attack performance, thereby more sharply highlighting the privacy risks users face in decoupled tool‑invocation frameworks.

\textbf{Insufficient Privacy‑Sensitive Tools.}
Our experiments with IntentMiner use open‑source datasets commonly used for evaluating tool retrieval methods.
However, these datasets include few tool calls involving privacy‑sensitive information.
For instance, in the ToolACE dataset, only 204 of 11,300 dialogues contain the keyword “health,” and some even refer to environmental rather than human health.
We believe IntentMiner should be further tested on datasets with more privacy‑sensitive tools, such as those providing health or legal advice.

\section*{Ethical Considerations}
\textbf{Non-Malicious Use and Defensive Purposes}
IntentMiner proposed in this work is not designed to acquire or disclose user sensitive information, but rather to advance user privacy protection.
Our ultimate goal is to reveal potential privacy risks within the MCP framework, thereby motivating practical defense strategies to enhance its overall security.

\textbf{Open-Source Data and Models}
All open‑source datasets and models used in our experiments are obtained from HuggingFace without modification. 
The commercial LLMs are accessed through their official APIs. Our use of open‑source resources fully complies with the corresponding data‑use agreements and open‑source licenses.

\textbf{Legal and Regulatory Compliance}
Our research uses legitimate and publicly available data that contain no sensitive personal information.
The purpose of this study is to identify privacy risks in the MCP framework, rather than to disclose any personal data.
Accordingly, this work complies with privacy and data‑protection regulations, including GDPR~\cite{gdpr2016}, CCPA~\cite{ccpa2018}, and the Cybersecurity Law~\cite{cybersecuritylaw2017}.

\textbf{Information About Use of AI Assistants}  
AI assistants were employed solely for auxiliary purposes during this research.  
Specifically, their use was limited to assisting with literature reading and improving the clarity of the manuscript.  
All code implementation, data analysis, and scientific writing were completed entirely by the authors without AI-generated content.





\bibliography{custom}

@article{zhao2023survey,
  title={A survey of large language models},
  author={Zhao, Wayne Xin and Zhou, Kun and Li, Junyi and Tang, Tianyi and Wang, Xiaolei and Hou, Yupeng and Min, Yingqian and Zhang, Beichen and Zhang, Junjie and Dong, Zican and others},
  journal={arXiv preprint arXiv:2303.18223},
  volume={1},
  number={2},
  year={2023}
}

@article{hadi2023survey,
  title={A survey on large language models: Applications, challenges, limitations, and practical usage},
  author={Hadi, Muhammad Usman and Qureshi, Rizwan and Shah, Abbas and Irfan, Muhammad and Zafar, Anas and Shaikh, Muhammad Bilal and Akhtar, Naveed and Wu, Jia and Mirjalili, Seyedali and others},
  journal={Authorea Preprints},
  year={2023},
  publisher={Authorea}
}

@article{chang2024survey,
  title={A survey on evaluation of large language models},
  author={Chang, Yupeng and Wang, Xu and Wang, Jindong and Wu, Yuan and Yang, Linyi and Zhu, Kaijie and Chen, Hao and Yi, Xiaoyuan and Wang, Cunxiang and Wang, Yidong and others},
  journal={ACM transactions on intelligent systems and technology},
  volume={15},
  number={3},
  pages={1--45},
  year={2024},
  publisher={ACM New York, NY}
}

@online{mcp2025,
  author   = {{Anthropic, PBC}},
  title    = {Model Context Protocol},
  year     = {2025},
  url      = {https://github.com/modelcontextprotocol},
  urldate  = {2025-09-25},
}

@misc{liu2024toolacewinningpointsllm,
      title={ToolACE: Winning the Points of LLM Function Calling}, 
      author={Weiwen Liu and Xu Huang and Xingshan Zeng and Xinlong Hao and Shuai Yu and Dexun Li and Shuai Wang and Weinan Gan and Zhengying Liu and Yuanqing Yu and Zezhong Wang and Yuxian Wang and Wu Ning and Yutai Hou and Bin Wang and Chuhan Wu and Xinzhi Wang and Yong Liu and Yasheng Wang and Duyu Tang and Dandan Tu and Lifeng Shang and Xin Jiang and Ruiming Tang and Defu Lian and Qun Liu and Enhong Chen},
      year={2024},
      eprint={2409.00920},
      archivePrefix={arXiv},
      primaryClass={cs.LG},
      url={https://arxiv.org/abs/2409.00920}, 
}

@article{song2020mpnet,
  title={Mpnet: Masked and permuted pre-training for language understanding},
  author={Song, Kaitao and Tan, Xu and Qin, Tao and Lu, Jianfeng and Liu, Tie-Yan},
  journal={Advances in neural information processing systems},
  volume={33},
  pages={16857--16867},
  year={2020}
}

@inproceedings{devlin2019bert,
  title={Bert: Pre-training of deep bidirectional transformers for language understanding},
  author={Devlin, Jacob and Chang, Ming-Wei and Lee, Kenton and Toutanova, Kristina},
  booktitle={Proceedings of the 2019 conference of the North American chapter of the association for computational linguistics: human language technologies, volume 1 (long and short papers)},
  pages={4171--4186},
  year={2019}
}

@misc{openai2024gpt4_1,
  title        = {GPT-4.1 Model Overview},
  author       = {{OpenAI}},
  year         = {2024},
  howpublished = {\url{https://platform.openai.com/docs/models/gpt-4.1}}
}

@misc{openai2025gpt5,
  title        = {GPT-5 Model Overview},
  author       = {{OpenAI}},
  year         = {2025},
  howpublished = {\url{https://platform.openai.com/docs/models/gpt-5}}
}

@misc{anthropic2024claude3,
  title        = {Claude 3 technical report},
  author       = {{Anthropic}},
  year         = {2024},
  institution  = {Anthropic PBC},
  howpublished = {\url{https://assets.anthropic.com/m/61e7d27f8c8f5919/original/Claude-3-Model-Card.pdf}}
}

@misc{anthropic2025claude4,
  title        = {Claude 4 technical report},
  author       = {{Anthropic}},
  year         = {2025},
  institution  = {Anthropic PBC},
  howpublished = {\url{https://www-cdn.anthropic.com/6be99a52cb68eb70eb9572b4cafad13df32ed995.pdf}}
}

@misc{google2025gemini25,
  title        = {Gemini 2.5 API and Model Documentation},
  author       = {{Google DeepMind}},
  year         = {2025},
  howpublished = {\url{https://ai.google.dev/gemini-api/docs}},
}

@misc{meta2024llama3,
  title        = {Open-source AI Models for Any Application: Llama 3},
  author       = {{Meta AI}},
  year         = {2024},
  howpublished = {\url{https://www.llama.com/models/llama-3/#models}},
}

@misc{deepseekai2024deepseekv3technicalreport,
      title={DeepSeek-V3 Technical Report}, 
      author={DeepSeek-AI},
      year={2024},
      eprint={2412.19437},
      archivePrefix={arXiv},
      primaryClass={cs.CL},
      url={https://arxiv.org/abs/2412.19437}, 
}

@misc{deepseekai2025deepseekr1incentivizingreasoningcapability,
      title={DeepSeek-R1: Incentivizing Reasoning Capability in LLMs via Reinforcement Learning}, 
      author={DeepSeek-AI},
      year={2025},
      eprint={2501.12948},
      archivePrefix={arXiv},
      primaryClass={cs.CL},
      url={https://arxiv.org/abs/2501.12948}, 
}

@article{qwen,
  title={Qwen Technical Report},
  author={Jinze Bai and Shuai Bai and Yunfei Chu and Zeyu Cui and Kai Dang and Xiaodong Deng and Yang Fan and Wenbin Ge and Yu Han and Fei Huang and Binyuan Hui and Luo Ji and Mei Li and Junyang Lin and Runji Lin and Dayiheng Liu and Gao Liu and Chengqiang Lu and Keming Lu and Jianxin Ma and Rui Men and Xingzhang Ren and Xuancheng Ren and Chuanqi Tan and Sinan Tan and Jianhong Tu and Peng Wang and Shijie Wang and Wei Wang and Shengguang Wu and Benfeng Xu and Jin Xu and An Yang and Hao Yang and Jian Yang and Shusheng Yang and Yang Yao and Bowen Yu and Hongyi Yuan and Zheng Yuan and Jianwei Zhang and Xingxuan Zhang and Yichang Zhang and Zhenru Zhang and Chang Zhou and Jingren Zhou and Xiaohuan Zhou and Tianhang Zhu},
  journal={arXiv preprint arXiv:2309.16609},
  year={2023},
  url={https://arxiv.org/abs/2309.16609}, 
}

@article{liu2023prompt,
  title={Prompt Injection attack against LLM-integrated Applications},
  author={Liu, Yi and Deng, Gelei and Li, Yuekang and others},
  journal={arXiv preprint arXiv:2306.05499},
  year={2023}
}

@article{greshake2023more,
  title={Not what you've signed up for: Compromising Real-World LLM-Integrated Applications with Indirect Prompt Injection},
  author={Greshake, Kai and others},
  journal={arXiv preprint arXiv:2302.12173},
  year={2023}
}

@inproceedings{fredrikson2015model,
  title={Model inversion attacks that exploit confidence information and basic countermeasures},
  author={Fredrikson, Matt and Jha, Somesh and Ristenpart, Thomas},
  booktitle={Proceedings of the 22nd ACM CCS},
  year={2015}
}

@misc{cablate2025mcp,
  author = {CabLate},
  title = {mcp-google-map: A powerful {Model Context Protocol (MCP)} server providing comprehensive {Google Maps API} integration with {LLM} processing capabilities},
  year = {2025},
  publisher = {GitHub},
  journal = {GitHub repository},
  howpublished = {\url{https://github.com/cablate/mcp-google-map}},
  note = {Accessed: 2025-12-14}
}

@inproceedings{pan2023privacy,
  title={Privacy Risks of General-Purpose Language Models},
  author={Pan, Xudong and others},
  booktitle={IEEE S\&P},
  year={2023}
}

@article{morris2023text,
  title={Text Embeddings Reveal (Almost) As Much As Text},
  author={Morris, John X and others},
  journal={arXiv preprint arXiv:2310.06816},
  year={2023}
}

@misc{zhan2024injecagentbenchmarkingindirectprompt,
      title={InjecAgent: Benchmarking Indirect Prompt Injections in Tool-Integrated Large Language Model Agents}, 
      author={Qiusi Zhan and Zhixiang Liang and Zifan Ying and Daniel Kang},
      year={2024},
      eprint={2403.02691},
      archivePrefix={arXiv},
      primaryClass={cs.CL},
      url={https://arxiv.org/abs/2403.02691}, 
}

@misc{wang2025mcptoxbenchmarktoolpoisoning,
      title={MCPTox: A Benchmark for Tool Poisoning Attack on Real-World MCP Servers}, 
      author={Zhiqiang Wang and Yichao Gao and Yanting Wang and Suyuan Liu and Haifeng Sun and Haoran Cheng and Guanquan Shi and Haohua Du and Xiangyang Li},
      year={2025},
      eprint={2508.14925},
      archivePrefix={arXiv},
      primaryClass={cs.CR},
      url={https://arxiv.org/abs/2508.14925}, 
}

@inproceedings{greshake2023not,
  title={Not what you've signed up for: Compromising Real-World {LLM}-Integrated Applications with Indirect Prompt Injection},
  author={Greshake, Kai and Abdelnabi, Sahar and Mishra, Shailesh and Endres, Christoph and Holz, Thorsten and Fritz, Mario},
  booktitle={Proceedings of the 16th ACM Workshop on Artificial Intelligence and Security},
  pages={79--90},
  year={2023}
}

@inproceedings{tramer2016stealing,
  title={Stealing Machine Learning Models via Prediction {APIs}},
  author={Tram{\`e}r, Florian and Zhang, Fan and Juels, Ari and Reiter, Michael K and Ristenpart, Thomas},
  booktitle={25th USENIX Security Symposium (USENIX Security 16)},
  pages={601--618},
  year={2016}
}

@inproceedings{shokri2017membership,
  title={Membership Inference Attacks Against Machine Learning Models},
  author={Shokri, Reza and Stronati, Marco and Song, Congzheng and Shmatikov, Vitaly},
  booktitle={2017 IEEE Symposium on Security and Privacy (SP)},
  pages={3--18},
  year={2017},
  organization={IEEE}
}

@inproceedings{carlini2021extracting,
  title={Extracting Training Data from Large Language Models},
  author={Carlini, Nicholas and Tramer, Florian and Wallace, Eric and Jagielski, Matthew and Herbert-Voss, Ariel and Lee, Katherine and Roberts, Adam and Brown, Tom and Song, Dawn and Erlingsson, Ulfar and others},
  booktitle={30th USENIX Security Symposium (USENIX Security 21)},
  pages={2633--2650},
  year={2021}
}

@misc{gdpr2016,
  title        = {Regulation (EU) 2016/679 of the European Parliament and of the Council of 27 April 2016 on the protection of natural persons with regard to the processing of personal data and on the free movement of such data (General Data Protection Regulation)},
  howpublished = {\url{https://eur-lex.europa.eu/eli/reg/2016/679/oj}},
  year         = {2016},
  institution  = {European Union},
  note         = {Official Journal of the European Union, L119, 1--88}
}

@misc{ccpa2018,
  title        = {California Consumer Privacy Act of 2018 (CCPA)},
  howpublished = {\url{https://leginfo.legislature.ca.gov/faces/codes_displayText.xhtml?division=3.&part=4.&lawCode=CIV&title=1.81.5}},
  year         = {2018},
  institution  = {State of California Legislature},
  note         = {California Civil Code, Title~1.81.5, Sections~1798.100--1798.199}
}

@misc{cybersecuritylaw2017,
  title        = {Cybersecurity Law of the People's Republic of China},
  howpublished = {\url{http://www.cac.gov.cn/2016-11/07/c_1119867116.htm}},
  year         = {2017},
  institution  = {National People's Congress of the People's Republic of China},
  note         = {Adopted on November 7, 2016; effective June 1, 2017}
}

\appendix

\section{Appendix}
\label{sec:appendix}

\subsection{Case Study}
\label{case_study}
\subsubsection{Medical Tool Call}

\begin{tcolorbox}[
  colback=gray!4,
  colframe=black!70,
  boxrule=0.8pt,
  arc=2pt,
  title=\textbf{User Query},
  breakable  
]

As part of my commitment to improve my heart health, I need to adjust my diet. Please recommend heart-healthy food items considering my condition of high blood pressure, borderline high cholesterol, and prediabetes. I prefer to include fish and vegetables, and avoid red meat and dairy. Also, could you generate a week-long diet plan based on these recommendations?
    
\end{tcolorbox}

\begin{tcolorbox}[
  colback=gray!4,
  colframe=black!70,
  boxrule=0.8pt,
  arc=2pt,
  title=\textbf{Tool Call},
  breakable  
]

\texttt{
Heart\_Healthy\_Food\_Recommender(user\_ \\
health\_condition = \{blood\_pressure \\ 
: High, cholesterol\_level: Border- \\ 
line high, blood\_sugar\_level: Pre- 
diabetes\}, dietary\_preferences = [ \\
fish, vegetables]);
} 

\texttt{
Heart\_Healthy\_Diet\_Planner(user\_ \\
health\_data = \{age: 45, weight: 75, \\
height: 165, medical\_conditions: [ \\
High blood pressure, Borderline \\
high cholesterol, Prediabetes]\}, \\
lifestyle = \{physical\_activity\_ \\
level: Lightly active, dietary\_ \\
habits: Pescatarian\});
}

\end{tcolorbox}

\begin{tcolorbox}[
  colback=gray!4,
  colframe=black!70,
  boxrule=0.8pt,
  arc=2pt,
  title=\textbf{Inverted Intent},
  breakable  
]

To actively manage high blood pressure, borderline high cholesterol, and prediabetes through dietary intervention.

\end{tcolorbox}

\subsubsection{Legal Tool Call}

\begin{tcolorbox}[
  colback=gray!4,
  colframe=black!70,
  boxrule=0.8pt,
  arc=2pt,
  title=\textbf{User Query},
  breakable  
]

I'm representing a client in a legal negotiation scheduled for next week on Thursday. We already had a negotiation this Monday and reached several agreements. Could you help me analyze the outcome of this Monday's negotiation based on the summary I provide and then prepare a strategy for the upcoming negotiation next week?

This Monday's negotiation summary: The client and the opposing party discussed terms for a new software license agreement. Agreements: 

1. Lowering the upfront cost of the license to \$10,000 reached in 2 hours. 

2. Extending the support period from 1 year to 3 years reached in 1.5 hours. 

Legal context for next week's negotiation: Intellectual property rights regarding software licenses. Client's requirements: 

1. Non-disclosure agreement (priority 1). 

2. Minor increase in annual maintenance cost (priority 2). 

3. Flexibility in user limits (priority 3).
    
\end{tcolorbox}

\begin{tcolorbox}[
  colback=gray!4,
  colframe=black!70,
  boxrule=0.8pt,
  arc=2pt,
  title=\textbf{Tool Call},
  breakable  
]

\texttt{
legal\_negotiation.analyze\_outcome( \\
negotiation\_summary = "The client \\
and the opposing party discussed \\ 
terms for a new software license \\
agreement.", agreements = [\{agree- \\
ment: Lowering the upfront cost \\
of the license to \$10,000, time\_ \\
to\_achieve: 2 hours\}, \{agreement: \\
Extending the support period from \\
1 year to 3 years, time\_to\_achieve: \\
1.5 hours\}], negotiation\_date = \\
Monday); 
} 

\texttt{
legal\_negotiation.prepare\_strategy( \\
legal\_context = "Intellectual pro- \\
perty rights regarding software li- \\
censes", client\_requirements = [\{ \\
requirement: Non-disclosure agree- \\
ment, priority: 1\}, \{requirement: \\
"Minor increase in annual main- \\
tenance cost", priority: 2\}, \{ \\
requirement: Flexibility in user \\
limits, priority: 3\}], negotiation\_ \\
date = Thursday);
}

\end{tcolorbox}

\begin{tcolorbox}[
  colback=gray!4,
  colframe=black!70,
  boxrule=0.8pt,
  arc=2pt,
  title=\textbf{Inverted Intent},
  breakable  
]

To develop a strategic plan for an upcoming legal negotiation focusing on intellectual property rights related to software licenses, incorporating client priorities.

\end{tcolorbox}

\subsubsection{Financial Tool Call}

\begin{tcolorbox}[
  colback=gray!4,
  colframe=black!70,
  boxrule=0.8pt,
  arc=2pt,
  title=\textbf{User Query},
  breakable  
]

I'm planning to invest in a bank and I need to check its financial performance before making a decision. I want to analyze its 1st Quarter performance report for this year. The bank's ID is 'BKR789'. After retrieving the data, please calculate the ROA and ROE ratios. Also, perform a credit risk evaluation based on the bank's current debt level of 5000000 and the credit history events list I provide: 
\texttt{
[\{eventDate: 15/ \\ 
02/2026 10:30, outcome: Positive\}, \\
\{eventDate: 13/04/2026 12:15, out- \\
come: Negative\}].
}
    
\end{tcolorbox}

\begin{tcolorbox}[
  colback=gray!4,
  colframe=black!70,
  boxrule=0.8pt,
  arc=2pt,
  title=\textbf{Tool Call},
  breakable  
]

\texttt{
BankPerformanceAnalysis.retrieve- \\
QuarterlyReport(bankDetails = \{ \\
bankId: BKR789, reportDate: 04- \\ 
01\}); 
} 

\texttt{
FinancialHealthCheck.computeRatios( \\
financialData = \{totalAssets: 18, \\
000,000, totalEquity: 5,000,000, \\
netEarnings: 2,000,000, period: \\
2026-04-01 00:00\});
}

\texttt{
RiskAssessment.evaluateCreditRisk( \\
riskParameters = \{creditHistory: [ \\
\{eventDate: 15/02/2026 10:30, out- \\
come: Positive\}, \{eventDate: 13/04/ \\
2026 12:15, outcome: Negative\}], \\
currentDebtLevel: 5,000,000\});
}

\end{tcolorbox}

\begin{tcolorbox}[
  colback=gray!4,
  colframe=black!70,
  boxrule=0.8pt,
  arc=2pt,
  title=\textbf{Inverted Intent},
  breakable  
]

To investigat Bank BKR789's financial metrics for potential investment or lending decisions.

\end{tcolorbox}

\subsection{The Prompt of IntentMiner}
\label{prompt:intent_miner}

\begin{tcolorbox}[
  colback=gray!4,
  colframe=black!70,
  boxrule=0.8pt,
  arc=2pt,
  title=\textbf{The Prompt of IntentMiner},
  breakable  
]

You are an intent inference system designed to analyze user behavior through their tool usage patterns. 
Your task is to infer the user's underlying intent by examining the tools they have called, the parameters they provided, and the results they received.

Your analysis should be thorough and consider:

\hspace*{0.5em} - The semantic meaning of tool names and descriptions

\hspace*{0.5em} - The specific parameters users provide when calling tools

\hspace*{0.5em} - The actual results returned by the tools

\hspace*{0.5em} - Patterns and relationships between multiple tool calls

\hspace*{0.5em} - Context clues from the combination of all available information

You must provide the most possible user intent, which should be specific, actionable, and grounded in the evidence from the tool calls.

Think step-by-step through the analysis process before concluding with your final intent predictions.

\section*{CONTEXT}
Analyze the following tool call information and infer the user's intent:

Tool Call Information: \$\{tools\_info\}

Above is a list of tool calls made by the user. Each tool call contains the following information:

\textbf{1. Tool Name}: The identifier / name of the tool

\textbf{2. Tool Description}: A detailed description of the tool's purpose and application

\textbf{3. Tool Schema}: 

\hspace*{0.5em} - \textbf{Required Field}: Parameters that must be provided when calling the tool
   
\hspace*{0.5em} - \textbf{Properties Field}: Detailed specification of all information returned by the tool, including their descriptions and data types
   
\textbf{4. Tool Call Statement}: The actual call made by the user with all provided parameters

\textbf{5. Tool Returned Result}: The full result returned by the tool after execution

\section*{OUTPUT}
Please follow this thought chain step-by-step to infer the user's intent:
    
\textbf{Step 1: Tool Purpose Analysis}
For each tool in the list, parse and understand:

\hspace*{0.5em} - What is the specific purpose of this tool based on its name?

\hspace*{0.5em} - What does the tool's description tell us about its intended use case?

\hspace*{0.5em} - What domain or problem space does this tool address?

\textbf{Step 2: Call Statement Analysis}
For each tool call statement, examine:

\hspace*{0.5em} - What parameters did the user provide in their request?

\hspace*{0.5em} - How do the user's provided parameters relate to the 'Required Field' in the 'Tool Schema'?

\hspace*{0.5em} - Based on the tool purpose analysis in Step 1, what do these specific parameter values suggest about the user's intent?

\textbf{Step 3: Returned Result Analysis}
For each tool's returned result:

\hspace*{0.5em} - What information did the tool return based on the user's parameters?

\hspace*{0.5em} - How do the result returned by the tool relate to the 'Properties Field' in the 'Tool Schema'?

\hspace*{0.5em} - Did the result provide the type of information that would support specific user intents?

\hspace*{0.5em} - How might the user intent to do use this returned information?

\textbf{Step 4: Intent Inference}
Based on the complete analysis above, return the most possible user intent as a Python list. 
The intent should be a clear, specific statement about what the user is trying to achieve.

\textbf{Output Format Requirements:}

\hspace*{0.5em} - Only output the final Python list.

\hspace*{0.5em} - Do not include any explanation, reasoning, or text outside of the list.

\hspace*{0.5em} - The output must strictly follow this format:

''' \\
intent = ["Most probable intent"] \\
'''
    
\end{tcolorbox}

\subsection{The Prompt of LLM Evaluators}
\label{prompt:evaluator}

\begin{tcolorbox}[
  colback=gray!4,
  colframe=black!70,
  boxrule=0.8pt,
  arc=2pt,
  title=\textbf{The Prompt of LLM Evaluator},
  breakable  
]

You are an assistant designed to evaluate intent alignment between an original user query and an inferred intent.

Your task is to determine whether the user's original intent effectively matches the inferred intent.

Follow these steps carefully:

1. Analyze the user’s original query to understand user's core intent

2. Determine if the purpose or desired action in the original query is similar with the inferred intent

3. Ignore minor wording differences, synonyms, or rephrasing - focus on whether the underlying intent is similar

4. Classify your evaluation using exactly one of these labels: \\
   - "Correct": The inferred intent accurately captures the user's original intent \\
   - "Incorrect": The inferred intent does not align with the user's original intent

\section*{CONTEXT}

User Query: \$\{user\_query\}
    
Inferred Intent: \$\{inferred\_intent\}

\section*{OUTPUT}

You should respond strictly following the specified output format:

''' \\
Result = Correct / Incorrect \\
''' \\
    
\end{tcolorbox}

\subsection{The Prompt of LLM-noCoT Inverter}
\label{prompt:llm_noCoT}

\begin{tcolorbox}[
  colback=gray!4,
  colframe=black!70,
  boxrule=0.8pt,
  arc=2pt,
  title=\textbf{The Prompt of LLM-noCoT Inverter},
  breakable  
]

You are an intent inference system designed to analyze user behavior through their tool usage patterns. 
Your task is to infer the user's underlying intent by examining the tools they have called, the parameters they provided, and the results they received.

You must provide the most possible user intent, which should be specific, actionable, and grounded in the evidence from the tool calls.

\section*{CONTEXT}

Analyze the following tool call information and infer the user's intent:

Available Tools Description: \$\{tools\_desc\}

Tool Call Information: \$\{tools\_info\}

\section*{OUTPUT}

\textbf{Output Format Requirements:}

\hspace*{0.5em} - Only output the final Python list.

\hspace*{0.5em} - Do not include any explanation, reasoning, or text outside of the list.

\hspace*{0.5em} - The output must strictly follow this format:

''' \\
intent = ["Most probable intent"] \\
'''
    
\end{tcolorbox}

\subsection{The Prompt of LLM-CoT Inverter}
\label{prompt:llm_CoT}

\begin{tcolorbox}[
  colback=gray!4,
  colframe=black!70,
  boxrule=0.8pt,
  arc=2pt,
  title=\textbf{The Prompt of LLM-CoT Inverter},
  breakable  
]

You are an intent inference system designed to analyze user behavior through their tool usage patterns. 
Your task is to infer the user's underlying intent by examining the tools they have called, the parameters they provided, and the results they received.

You must provide the most possible user intent, which should be specific, actionable, and grounded in the evidence from the tool calls.

\section*{CONTEXT}

Analyze the following tool call information and infer the user's intent:

Available Tools Description: \$\{tools\_desc\}

Tool Call Information: \$\{tools\_info\}

\section*{OUTPUT}

Please follow this thought chain step-by-step to infer the user's intent:

\textbf{Step 1: Tool Description Analysis} Analyze the purpose of the invoked tool based on the information provided in the Available Tools Description.

\textbf{Step 2: Call Statement Analysis} Based on the Tool Call Information, extract and analyze the parameters supplied during the tool call.

\textbf{Step 3: Returned Result Analysis} Based on the Tool Call Information, extract and analyze the results produced by the tool execution.

\textbf{Step 4: Intent Inference} Based on the complete analysis above, return the most possible user intent as a Python list. 
The intent should be a clear, specific statement about what the user is trying to achieve.

\textbf{Output Format Requirements:}

\hspace*{0.5em} - Only output the final Python list.

\hspace*{0.5em} - Do not include any explanation, reasoning, or text outside of the list.

\hspace*{0.5em} - The output must strictly follow this format:

''' \\
intent = ["Most probable intent"] \\
'''
    
\end{tcolorbox}

\end{document}